\documentclass[epj,nopacs]{svjour}
\usepackage{amssymb}
\usepackage{graphicx}
\usepackage{dcolumn}
\usepackage{bm}
\usepackage{color}
\newcommand{\be}{\begin{equation}}
\newcommand{\ee}{\end{equation}}

\newcommand{\One}{1\hspace*{-3.65pt}1}
\newcommand{\rse}{\mathcal{R}}
\newcommand{\bes}[1]{j^{#1}_{L_{#1}}}
\newcommand{\han}[1]{h^{(1)#1}_{L_{#1}}}

\newcommand{\tmat}[2]{T_{#1#2}^{(L_{#1},L_{#2})}}
\newcommand{\ttpo}{$2\,{}^{3\!}P_1$}
\newcommand{\otpo}{$1\,{}^{3\!}P_1$}
\newcommand{\jpsi}{J\!/\!\psi}
\begin{document}
\title{Delicate interplay between the $D^0D^{\ast0}$, $\rho^0J\!/\!\psi$,
and $\omega J\!/\!\psi$ channels \\ in the $X(3872)$ resonance}
\author{Susana Coito\inst{1} \and George Rupp\inst{1} \and
       Eef van Beveren\inst{2}
}                     
\institute{
Centro de F\'{\i}sica das Interac\c{c}\~{o}es Fundamentais,
Instituto Superior T\'{e}cnico, Technical University of Lisbon,
P-1049-001 Lisbon, Portugal
\and
Centro de F\'{\i}sica Computacional,
Departamento de F\'{\i}sica, Universidade de Coimbra,
P-3004-516 Coimbra, Portugal
}
\date{Received: date / Revised version: date}
%
\abstract{
The nature of the $X(3872)$ enhancement is analysed in the
framework of the Resonance-Spectrum Expansion, by studying it as a regular
$J^{PC}=1^{++}$ charmonium state, though strongly influenced and shifted by
open-charm decay channels.
The observed but Okubo-Zweig-Iizuka-forbidden
$\rho^0J\!/\!\psi$ and $\omega J\!/\!\psi$ channels are coupled as well, but
effectively smeared out by using complex $\rho^0$ and $\omega$ masses,
in order to account for their physical widths, followed by a rigorous
algebraic procedure to restore unitarity.
A very delicate interplay between 
the $D^0D^{\ast0}$, $\rho^0J\!/\!\psi$, and $\omega J\!/\!\psi$
channels is observed. The data clearly
suggest that the $X(3872)$ is a very narrow axial-vector $c\bar{c}$ resonance,
with a pole at or slightly below the $D^0D^{\ast0}$ threshold.
%
} 

\maketitle

\section{Introduction}
The $X(3872)$ charmonium-like state was discovered in 2003 by the Belle
Collaboration \cite{PRL91p262001}, as a $\pi^+\pi^- J\!/\!\psi$ enhancement
in the decay $B^{\pm} \rightarrow K^{\pm} \pi^+\pi^- J\!/\!\psi$. The
same structure was then observed, again in $\pi^+\pi^- J\!/\!\psi$, by 
CDF II \cite{PRL93p072001}, D0 \cite{PRL93p162002}, and BABAR
\cite{PRD71p071103}. Moreover, CDF \cite{PRL96p102002} showed that the
$\pi^+\pi^-$ mass distribution favors decays via a $\rho^0$ resonance,
implying positive $C$-parity for the $X(3872)$. The $X(3872)$ has also
been observed in the $\bar{D}^0D^0\pi^0$ and $\bar{D}^{\ast0}D^{\ast0}$
channels, by Belle \cite{PRL97p162002} and BABAR \cite{PRD77p011102},
respectively.
More recently, CDF \cite{PRL103p152001}
measured the $X(3872)$ mass with even higher precision, viz.\
$3871.61\pm0.16\pm0.19$~MeV, with a width fixed at $1.34\pm0.64$~MeV,
while BABAR \cite{PRD82p011101} presented evidence for the long-awaited
$\omega J\!/\!\psi$ decay mode
(also see Ref.\ \cite{HEPEX0505037}),
and a surprising preference for the
$2^{-+}$ assignment. We shall come back to the latter claim below.
The $X(3872)$ resonance is listed in the 2010 PDG tables
\cite{JPG37p075021}, with a mass of $3871.56\pm0.22$ MeV, a width
$<\!2.3$ MeV, and $1^{++}$ or $2^{-+}$ quantum numbers
\cite{JPG37p075021,PRL98p132002}.

On the theoretical side, the first to foresee a narrow $1^{++}$ state
close to the $DD^\ast$ threshold was T\"{o}rnqvist \cite{ZPC61p525},
arguing on the basis of strongly attractive one-pion exchange for
$S$-wave meson-meson systems, which he called deusons. For
further
molecular descriptions and studies, see Ref.~\cite{molecular}, as well as
the reviews by Swanson \cite{PR429p243} and Klempt \& Zaitsev \cite{PR454p1}.
In Ref.~\cite{exotics}, a few exotic model descriptions can be found, such
as a hybrid or a tetraquark; also see the reviews \cite{PR429p243,PR454p1}.
For further reading, we  recommend the very instructive analyses by
Bugg \cite{PRD71p016006} and Kalashnikova \& Nefediev \cite{PRD80p074004}.

Much more in the spirit of our own calculation is the coupled-channel analysis
by Danilkin and Simonov \cite{PRD81p074027}, which studies resonances and
level shifts of conventional charmonium states due to the most important open
and closed decay channels. We shall come back to their results below.

As said above, according to the PDG \cite{JPG37p075021}, the $X(3872)$ is
either a $1^{++}$ or a $2^{-+}$ state, which implies $2\,{}^{3\!}P_1$ or
$1\,{}^{1\!}D_2$, as other radial excitations would be much too far off (see
e.g.\ Ref.~\cite{PRD32p189}). In the present paper, we only study the $1^{++}$
scenario, despite the conclusion by BABAR \cite{PRD82p011101}, from the
$\omega J\!/\!\psi$ mode, that $2^{-+}$ is more likely. However, the latter
assignment
appears to be
at odds with radiative-transition data \cite{1007.4541}.
For a further discussion of electromagnetic decays, see e.g.\ the molecular
description of Ref.~\cite{PRD79p094013}.
But more importantly,
in all charmonium models we know of, the
$1\,{}^{1\!}D_2$ $c\bar{c}$ state lies well below 3.872~GeV, i.e., in the range
3.79--3.84~GeV
(see e.g.\ Ref.~\cite{1011.6124}).
Our own \em bare \em \/$1\,{}^{1\!}D_2$ state comes out at
3.79~GeV, just as the corresponding single-channel state in
Ref.~\cite{PRD81p074027}. Now, the crucial point is that loops from closed
meson-meson channels are \em always \em \/attractive \cite{ZPC19p275}.
Hence, since
$DD^\ast$ at 3.872--3.880~GeV is the lowest Okubo-Zweig-Iizuka-allowed (OZIA)
channel that couples to a $1\,{}^{1\!}D_2$ $c\bar{c}$ state,
the coupled-channel mass shift will inexorably be further downwards
(also see Ref.\ \cite{1011.6124}).

The goal of the present paper is to show that the mass and width of the
$X(3872)$, as well as the corresponding observed amplitudes in the
$D^0D^{\ast0}$, $\rho^0J\!/\!\psi$, and $\omega J\!/\!\psi$ channels,
are compatible with a description in terms of a regular \ttpo\ 
charmonium state, though mass-shifted and unitarised via open and
closed decay channels.

\section{Resonance-Spectrum Expansion}
Sticking to the $1^{++}$ scenario, we employ again the Reso\-nance-Spectrum
Expansion (RSE) \cite{AOP324p1620}, in order to couple one $c\bar{c}$ channel,
with $l_c=1$, to several OZIA pseudo\-scalar-vector (PV) and vector-vector
(VV) channels, just as in our recent preliminary study \cite{1005.2486} of the
$X(3872)$. However, we now also couple the 
OZI-forbidden (OZIF) $\rho^0\jpsi$ and $\omega\jpsi$ channels,
to account for the bulk of the observed $\pi^+\pi^-\jpsi$
and $\pi^+\pi^-\pi^0\jpsi$ decays, respectively.
Although
the former channel is
isospin breaking 
as well,
the extreme closeness of its
central threshold at 3872.4~MeV to the $X(3872)$ structure makes it absolutely
nonnegligible, despite a very small expected coupling. A complication, though,
is the large $\rho$ width, which does not allow the $\rho^0\jpsi$ channel to
be described through a sharp threshold. 
Effects in the $X(3872)$ from non-zero $\rho$ and $\omega$ widths 
were already estimated in Ref.~\cite{PRD76p034007}.
We tackle this problem by taking a complex mass for the $\rho$, from its pole
position \cite{EPJA44p425}, and then apply a novel, empirical yet
rigorous, unitarisation procedure to the $S$-matrix, derived in
Appendix~\ref{unitarization}. The analyticity and causality implications of
complex masses in asymptotic states were already studied a long time ago
\cite{NPB12p281}.

For consistency, we apply the same procedure to the $\omega$ meson, despite
the fact that its width is a factor 17.5 smaller than that of the
$\rho$. Nevertheless, the $\omega$ width of about 8.5~MeV is very close
to the energy difference between the $\omega\jpsi$ and $D^0D^{\ast0}$
thresholds, and therefore not negligible. Finally, we shall neglect the
unknown small ($<2.1$ MeV \cite{JPG37p075021}) $D^{\ast0}$ width
\cite{PRD80p074004}, because of the relatively large error bars on the
$D^0D^{\ast0}$ data, though this width may have some influence on the 
precise $X(3872)$ pole position.
Nevertheless, reasonable estimates of the $D^{\ast0}$ width yield values
clearly smaller than 100~keV \cite{PRD76p094028}, so that its effect should
be largely negligible as compared to that of the $\rho$ and $\omega$ widths.

Let us now proceed with our RSE calculation of a bare $2\,{}^{3\!}P_1$
(with $n\!=\!1$, $J\!=\!1$, $L\!=\!1$, $S\!=\!1$) $c\bar{c}$ state,  coupled
to a number of MM channels. The resulting closed-form $T$-matrix is given in
Appendix~\ref{tmatrix}. In Table~\ref{MM} we list the
\begin{table}[ht]
\caption{Included meson-meson channels, with thresholds and ground-state
couplings. For simplicity, we omit the bars over the anti-charm mesons;
also note that $D^\ast D^\ast$ stands for the corresponding mass-averaged
charged and uncharged channels.}
\centering
\begin{tabular}{l|c|c|c}
\hline \hline &&& \\[-9pt]
Channel & $\left(g^i_{(l_c=1,n=0)}\right)^2$ & $\;L\;$ &
Threshold (MeV) \\[2pt] 
\hline &&& \\[-9pt]
$\rho^0 J/\psi$	& variable  & 0 & $3872.406 - i\,74.7$\\
$\omega J/\psi$	& variable  & 0 & $3879.566 - i\,4.25$\\
\hline &&& \\[-9pt]
$D^0D^{\ast0}$	& 1/54 & 0 & 3871.81\\ 
$D^0D^{\ast0}$ 	& 5/216 & 2 & 3871.81\\ 
$D^{\pm}D^{*\mp}$ & 1/54 & 0 & 3879.84\\
$D^{\pm}D^{*\mp}$ & 5/216 & 2 & 3879.84\\
$D^{*}D^{*}$ & 5/36 & 2 & 4017.24\\ 
$D_s^{\pm}D_s^{*\mp}$& 1/54 & 0 & 4080.77\\ 
$D_s^{\pm}D_s^{*\mp}$& 5/216 & 2 & 4080.77\\
\hline \hline
\end{tabular}
\label{MM}
\end{table}
considered PV and VV channels, including
$\rho^0\jpsi$ and $\omega\jpsi$.
Besides
the latter two OZIF channels and the also observed OZIA $D^0D^{\ast0}$
channel, we furthermore account for the OZIA PV and VV channels
$D^\pm D^{\ast\mp}$, $D^\ast D^\ast$, and $D_s D_s^\ast$, whose influence
on the $X(3872)$ pole position is not negligible, in spite of
being closed channels. The $D_s^\ast D_s^\ast$ channel, with threshold
about 350~MeV above the $X(3872)$ mass, we do not include.

The relative couplings
of the OZIA channels have been computed with the formalism of
Ref.~\cite{ZPC21p291}.
Couplings calculated in the latter scheme for ground-state mesons generally
coincide with the usual recouplings of spin, isospin, and orbital angular
momentum. Moreover, for excited stat\-es the formalism yields clear
predictions as well, contrary to other appoaches.
In Table~\ref{MM}, the squares of the ground-state
($n\!=\!0$) couplings are given, which have to be multiplied by $(n+1)/4^n$
for the $S$-wave PV channels, and by $(2n/5+1)/4^n$ for the others, so as to
obtain the couplings in the RSE sum of Eq.~(\ref{rse}). Also note that the
two (closed) $D^\ast D^\ast$ channels have been lumped together, with their
average threshold value and sum of squared couplings. As for the
couplings
of the
$\rho^0\jpsi$ and $\omega\jpsi$ channels,
the formalism of Ref.~\cite{ZPC21p291}, based on
OZIA decay via ${}^{3\!}P_0$ quark-pair creation, cannot make any prediction.
However, we know from experiment that the
couplings of OZIF channels are considerably smaller than those of OZIA
channels. Moreover, isospin-breaking channels are even further suppressed.
Thus, in the following we shall employ the values
$g_{\rho^0\jpsi}=0.07\times g_{D^0D^{\ast0}}$ and
$g_{\omega\jpsi}=0.21\times g_{D^0D^{\ast0}}$, which correspond to effective
relative strengths of 0.49\% and 4.41\%, respectively, which seem reasonable
to us. These values may also be compared to the corresponding relative
probabilites of about 0.65\% ($\approx0.006/0.92$) and 4.5\%
($\approx0.041/0.92$), respectively, employed in Ref.~\cite{PRD79p094013}.
Furthermore, we shall also test coupling values twice as large,
namely $g_{\rho^0\jpsi}=0.14\times g_{D^0D^{\ast0}}$ and
$g_{\omega\jpsi}=0.42\times g_{D^0D^{\ast0}}$.
Note that our coupling for the isospin-breaking channel $\rho^0\jpsi$ is
also in rough agreement with estimates from the rate of the observed
\cite{JPG37p075021} isospin-violating $\omega\to\pi^+\pi^-$ decay, which
amounts to about 1.5\% of the total width.
Another difference
between OZIA and OZIF channels is the average distance $r_i$ (see
Eqs.~(\ref{tmat},\ref{omega})) at which a light $q\bar{q}$ pair is created
before decay, which in the OZIA case we believe to take place in the core
region and in the OZIF case more in the periphery. Thus, we
employ
a larger
value for $r_1\equiv r_{\rho^0\jpsi}$
$=r_{\omega\jpsi}$
than for $r_0$, the single radius used
for all OZIA channels. Concretely, we take $r_0=2$~GeV$^{-1}\simeq0.4$~fm
and
$r_1=3$~GeV$^{-1}\simeq0.6$~fm,
while we also test the case $r_1=r_0$.

For the bare $c\bar{c}$ energy levels $E_n^{(l_c)}$ in the RSE sum of
Eq.~(\ref{rse}), we take an equidistant harmonic oscillator (HO), as in
previous work (see e.g.\ Ref.~\cite{PRD80p094011}). The corresponding charm
quark mass and HO frequency are once again kept fixed at the values
$m_c=1562$~MeV and $\omega=190$~MeV. The only parameter we adjust freely is
the overall coupling constant $\lambda$ in Eqs.~(\ref{tmat},\ref{omega}), which
is tuned to move the bare \ttpo\ state from 3979~MeV down to 
the $D^0D^{\ast0}$ threshold, requiring a $\lambda$ value of the order
of 3, i.e., not far from the values used in e.g.\
Refs.~\cite{PRD80p094011,PLB641p265}. At the
same time, the bare \otpo\ state shifts from 3599 MeV down to 
about 3.55 GeV, though depending quite sensitively on the precise form of the
used subthreshold suppression of closed channels \cite{PRD80p094011}. Anyhow,
for the purpose of the present study, an accurate reproduction of the
$\chi_{c1}(1P)$ mass of 3511 MeV is not very relevant.

\section{$X(3872)$ poles and amplitudes vs.\ data}
\begin{figure}[htb]
\begin{center}
\resizebox{!}{390pt}{\includegraphics{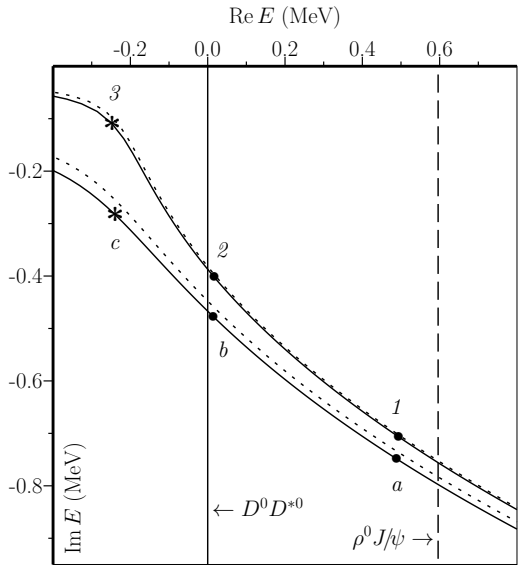}}
\mbox{} \\[-175pt]
\caption{Pole trajectories for $r_1$=3.0 GeV$^{-1}$ (solid curves)
and
$r_1$=2.0 GeV$^{-1}$ (dotted curves);
$g_{\rho^0\jpsi}/g_{D^0D^{\ast0}}=0.07$ and
$g_{\omega\jpsi}/g_{D^0D^{\ast0}}=0.21$ (upper two curves),
$g_{\rho^0\jpsi}/g_{D^0D^{\ast0}}=0.14$ and
$g_{\omega\jpsi}/g_{D^0D^{\ast0}}=0.42$ (lower two curves). Note that the
CM
energy
$E$
is relative to the $D^0D^{\ast0}$ threshold in all figures.
Also see Table~\ref{poles}.}
\label{trajectories}
\end{center}
\end{figure}

In Table~\ref{poles}, we give some pole positions in the vicinity of the
\begin{table}[h]
\caption{Pole positions of the dots and stars in Fig.~\ref{trajectories}.
In all cases, $r_1=3.0$ GeV$^{-1}$. Note that the OZIF couplings
$\tilde{g}_{\rho^0\jpsi}$ and $\tilde{g}_{\omega\jpsi}$ are given relative
to the coupling of the OZIA $D^0D^{\ast0}$ channel.}
\centering
\begin{tabular}{c|c|c|c|c}
\hline \hline &&&& \\[-9pt]
Label & $\lambda$ & $\tilde{g}_{\rho^0\jpsi}$
& $\tilde{g}_{\omega\jpsi}$ & Pole (MeV)\\[1pt] 
\hline &&&& \\[-9pt]
\emph{1} & 3.028 & 0.07 & 0.21 & $3872.30 - i\,0.71$\\
\emph{2} & 3.066 & 0.07 & 0.21 & $3871.83 - i\,0.40$\\ 
\emph{3} & 3.083 & 0.07 & 0.21 & $3871.56 - i\,0.11$\\ 
\emph{a} & 2.981 & 0.14 & 0.42 & $3872.30 - i\,0.75$\\
\emph{b} & 3.017 & 0.14 & 0.42 & $3871.82 - i\,0.48$\\
\emph{c} & 3.033 & 0.14 & 0.42 & $3871.57 - i\,0.28$\\ 
\hline \hline
\end{tabular}
\label{poles}
\end{table}
$D^0D^{\ast0}$ and $\rho^0\jpsi$ thresholds, with the chosen  values
of $\lambda$ and $r_1$. In Fig.~\ref{trajectories}, third-sheet pole
trajectories
in the complex energy plane (relative to the $D^0D^{\ast0}$ threshold)
are plotted, as a function of $\lambda$,
with the pole positions of Table~\ref{poles} marked by
bullets and stars.
The solid curves represent the case $r_1=3.0$~GeV$^{-1}$, while the dotted
ones stand for $r_1=2.0$~GeV$^{-1}$, showing little sensitivity to the
precise decay radius.
Figure~\ref{trajectories} shows that the $X(3872)$ resonance pole may come out
below the $D^0D^{\ast0}$ threshold
with a nonvanishing width, which is moreover of the right order of magnitude,
viz.\ $<\!1$~MeV.
The recent CDF \cite{PRL103p152001} mass determination of the $X(3872)$ might
suggest that the
pole positions '3' or `c' (see Table~\ref{poles} and Fig.~\ref{trajectories})
are favored.
However, one should realise that the differences amount to mere fractions of
an MeV,
while experimental uncertainties are at least of the same order.

\begin{figure}[h]
\begin{center}
\resizebox{!}{390pt}{\includegraphics{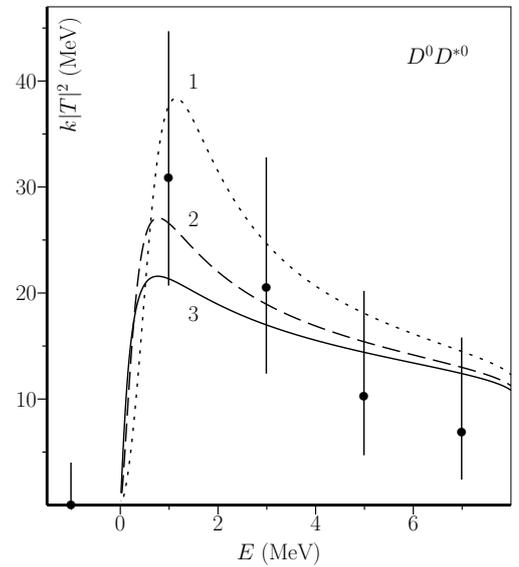}}
\mbox{} \\[-180pt]
\caption{$D^0D^{\ast0}$
elastic
amplitude for poles 1, 2, 3 in
Table~\ref{poles} and Fig.~\ref{trajectories}; arbitrarily normalised
data are from Ref.~\cite{0810.0358}.
Elastic $T$-matrix elements follow
from Eqs.~(\ref{tmat}--\ref{rse}); $k$ is on-shell relative momentum.}
\label{DD123}
\end{center}
\end{figure}

\begin{figure}[h]
\begin{center}
\resizebox{!}{390pt}{\includegraphics{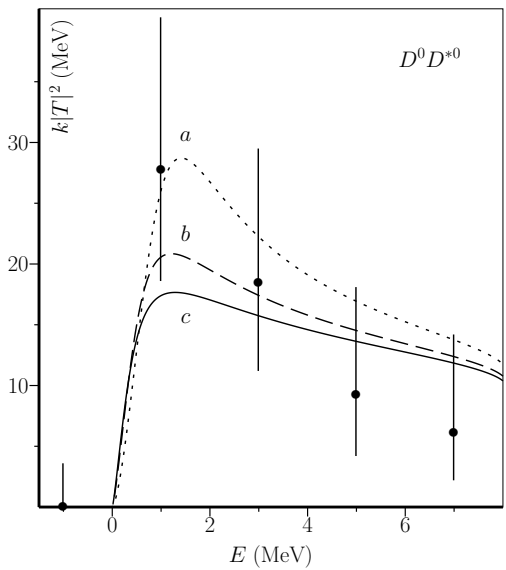}}
\mbox{} \\[-180pt]
\caption{As in Fig.~\ref{DD123}, but now for poles a, b, c.}
\label{DDabc}
\end{center}
\end{figure}

Now we compare the corresponding $D^0D^{\ast0}$ amplitudes to Belle
\cite{0810.0358} data , for the six cases labeled `1, 2, 3', and 'a, b, c' in
Table~\ref{poles} and Fig.~\ref{trajectories}. The results are depicted
in Figs.~\ref{DD123} and \ref{DDabc}, respectively. Note that
we allow for an arbitrary normalisation of the data, which is inevitable as
we are dealing with production data, 
which cannot be directly
compared with our scattering amplitudes,
also because of the finite experimental mass bins.
From these figures we see that the best agreement with data is
obtained in case `2', though 5 out of the 6 curves pass through all error
bars. Nevertheless, in view of the large errors, one should be very cautious
in drawing definite conclusions on the precise pole position as well as the
preferred OZIF couplings $g_{\rho^0\jpsi}$ and $g_{\omega\jpsi}$.

\begin{figure}
\begin{tabular}{cc}
\hspace*{-22pt}
\resizebox{!}{230pt}{\includegraphics{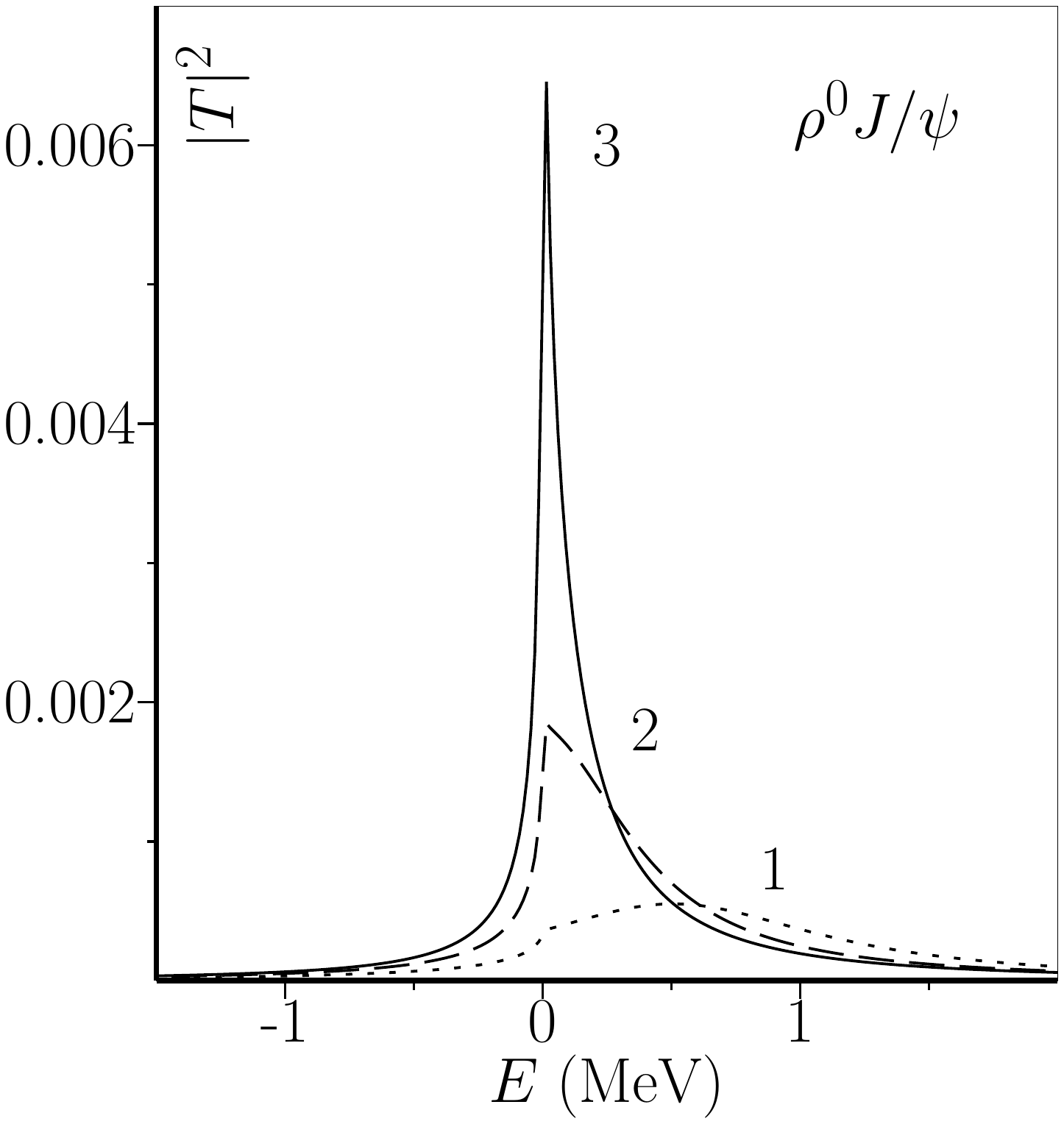}}
&
\hspace*{-64pt}
\resizebox{!}{230pt}{\includegraphics{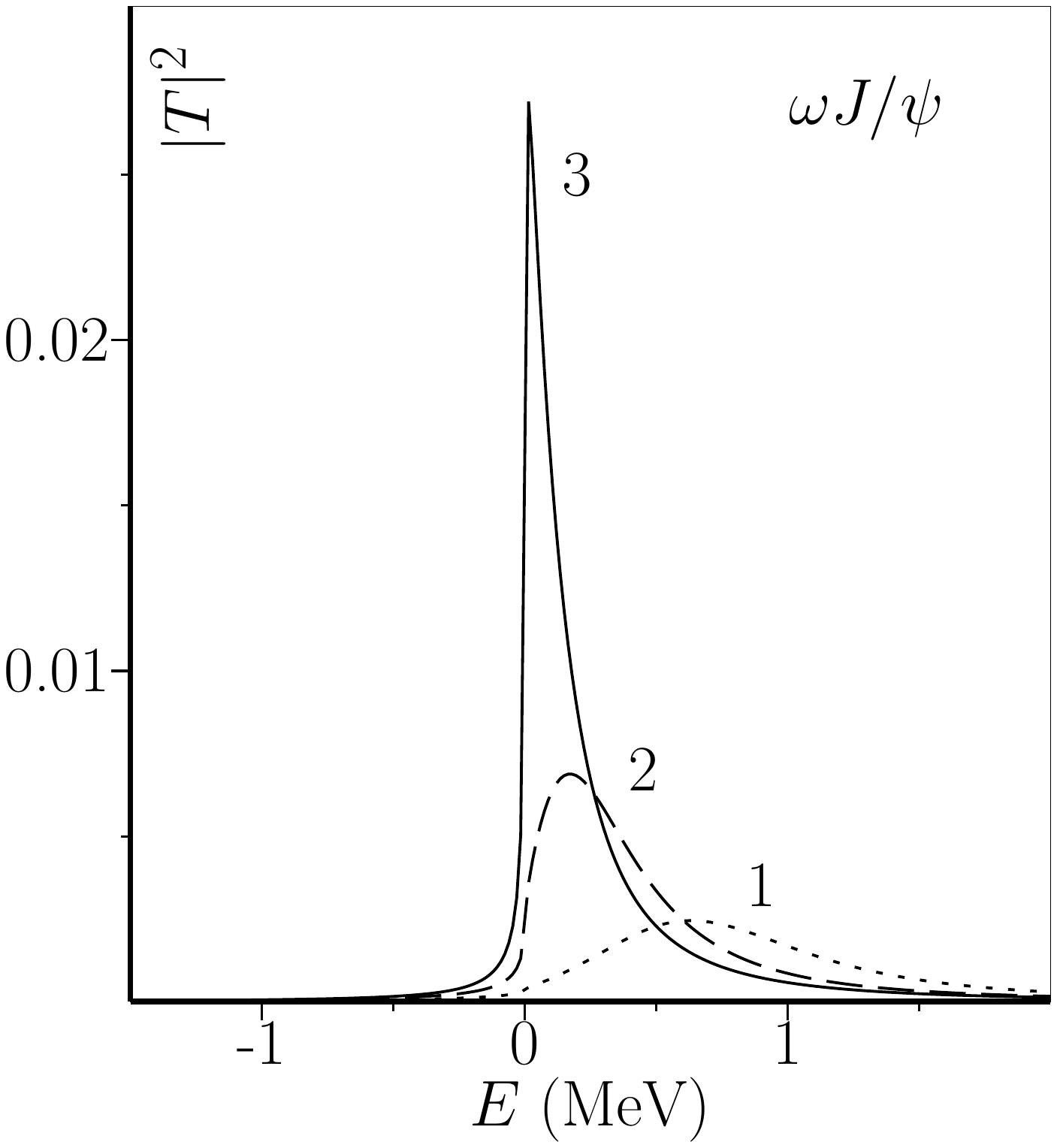}}
\end{tabular}
\mbox{ } \\[-110pt]
\caption{
$\rho^0\jpsi$ (left) and $\omega\jpsi$ (right) elastic amplitudes for
poles 1, 2, 3.
Also see
Fig.~\ref{trajectories} and
Table~\ref{poles}.}
\label{tsqrhoomeg123}
\end{figure}

Next we show, in Fig.~\ref{tsqrhoomeg123}, the elastic amplitudes in the
$\rho^0\jpsi$ and $\omega\jpsi$ channels, corresponding to the pole positions
1, 2, 3, i.e., for the smaller values of the OZIF couplings. We see that
both amplitudes are very
sensitive to the precise pole position, which
is logical, as the
OZIF channels couple
much more weakly to $c\bar{c}$ than $D^0D^{\ast0}$, so that the latter channel
will strongly deplete the former
ones,
as soon as it acquires some phase space. This is in line with our
analysis in e.g.\ Ref.~\cite{1005.1010}. Also note the strongly
cusp-like structure of the amplitude in the cases
2 and 3 for $\rho^0\jpsi$, and 3 for $\omega\jpsi$,
which is a manifestation of the depletion due to the opening of the
$D^0D^{\ast0}$ channel. Such a cusp makes the experimental determination
of the $X(3872)$ width very difficult.

\begin{figure}
\mbox{ } \\[5pt]
\begin{tabular}{cc}
\hspace*{-22pt}
\resizebox{!}{230pt}{\includegraphics{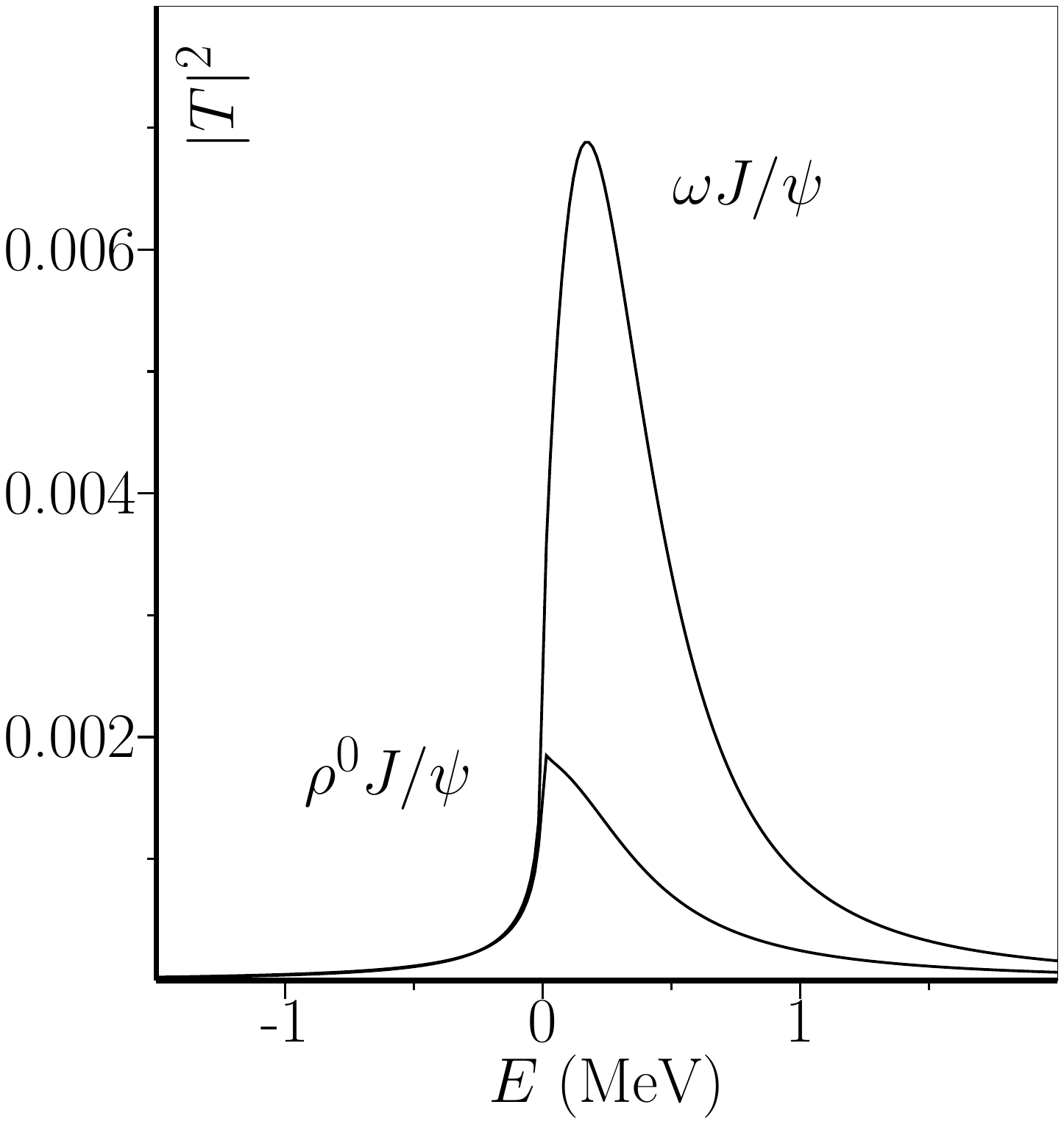}}
&
\hspace*{-64pt}
\resizebox{!}{230pt}{\includegraphics{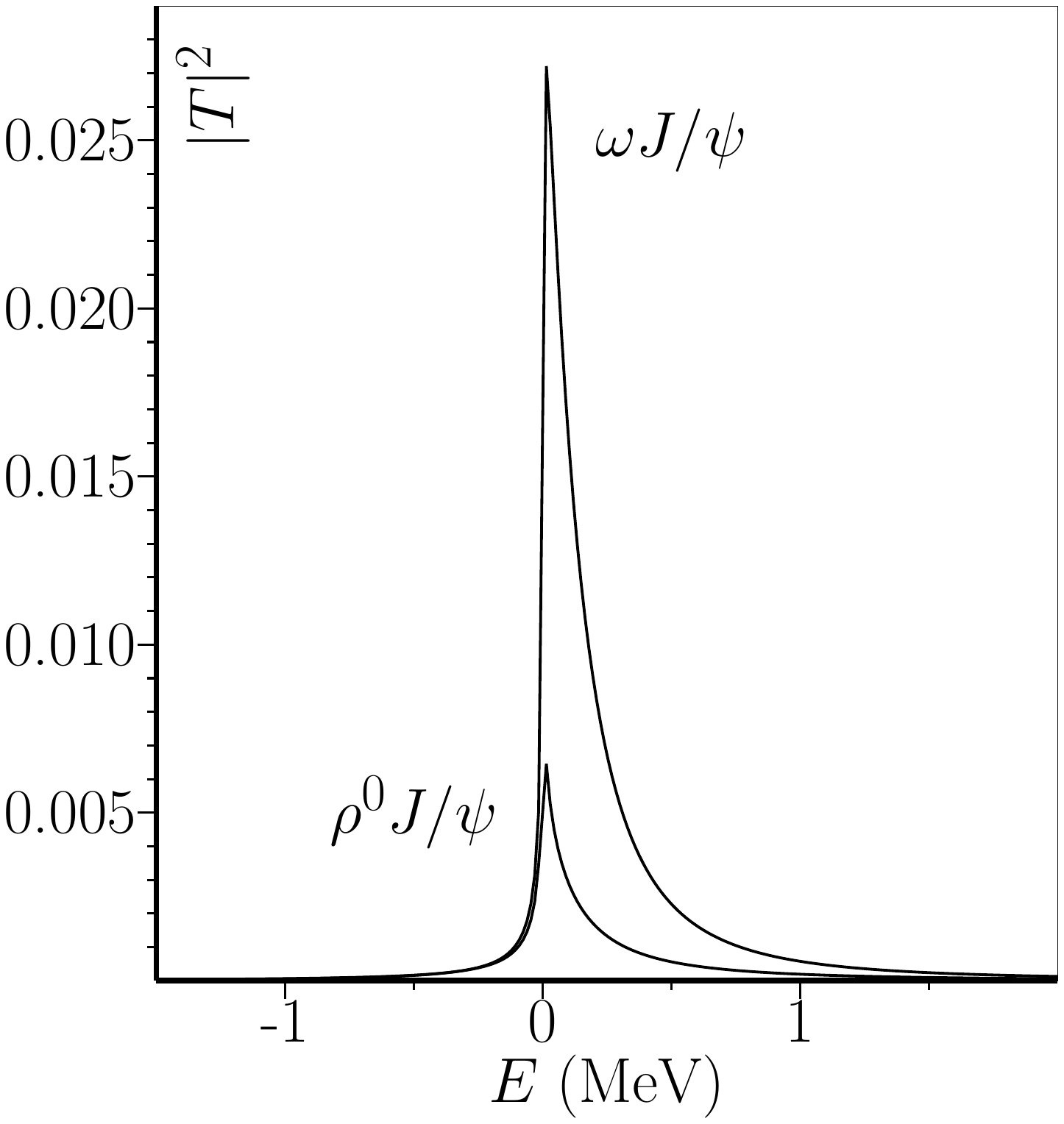}}
\end{tabular}
\mbox{ } \\[-110pt]
\caption{
$\rho^0\jpsi$ and $\omega\jpsi$ elastic amplitudes, for
poles 2 (left) and 3 (right).
Also see
Fig.~\ref{trajectories} and
Table~\ref{poles}.}
\label{tsqomegrho23}
\end{figure}

In Fig.~\ref{tsqomegrho23} we take a closer look at the $\omega\jpsi$ and
$\rho^0\jpsi$ amplitudes, in particular how they compare to one another. Now,
the effective strength of the $\omega\jpsi$ elastic $T$-matrix element is
9 times that of $\rho^0\jpsi$, as its coupling has been chosen 3 times as
large (see Table~\ref{poles} and Eqs.~(\ref{tmat}--\ref{rse})). For the
corresponding square amplitudes plotted in Fig.~\ref{tsqomegrho23}, this
amounts to a factor as large as 81. However, the central $\omega\jpsi$
threshold lies more than 7~MeV above that of $\rho^0\jpsi$, while the full
$\omega$ width is only 8.49~MeV. On the other hand, the central $\rho^0\jpsi$
threshold lies much closer to $D^0D^{\ast0}$, while the large physical $\rho$
width strongly boosts the associated amplitude, as demonstrated below. 
Qualitative arguments in agreement with our calculation were already presented
in Ref.~\cite{PRD80p014003}.
These effects make
the maximum $\omega\jpsi$ square amplitude to be only a factor 3.5--4 larger
than that of $\rho^0\jpsi$, both in case 2 and 3, as can be read off from
Fig.~\ref{tsqomegrho23}. Moreover, at the precise energy of the respective
pole position, the two amplitudes are almost equal in size. Therefore, the
observed branching ratio
$\mathcal{B}(X(3872)\to\omega\jpsi)/
\mathcal{B}(X(3872)\to\pi^+\pi^-\jpsi)\sim1$ \cite{PRD82p011101,HEPEX0505037}
is compatible with the present model calculation.

\begin{figure}
\begin{tabular}{cc}
\hspace*{-30pt}
\resizebox{!}{230pt}{\includegraphics{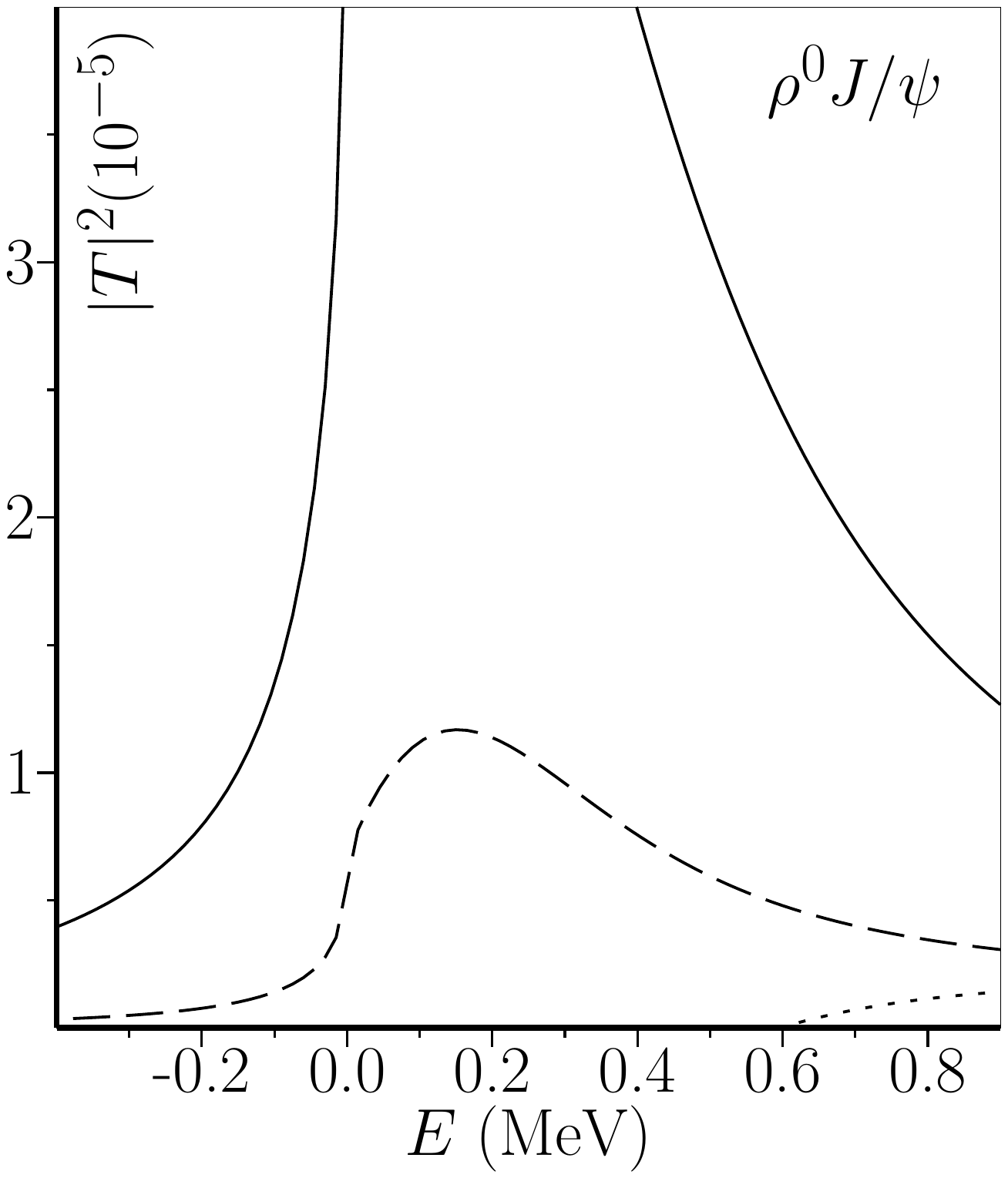}}
&
\hspace*{-60pt}
\resizebox{!}{230pt}{\includegraphics{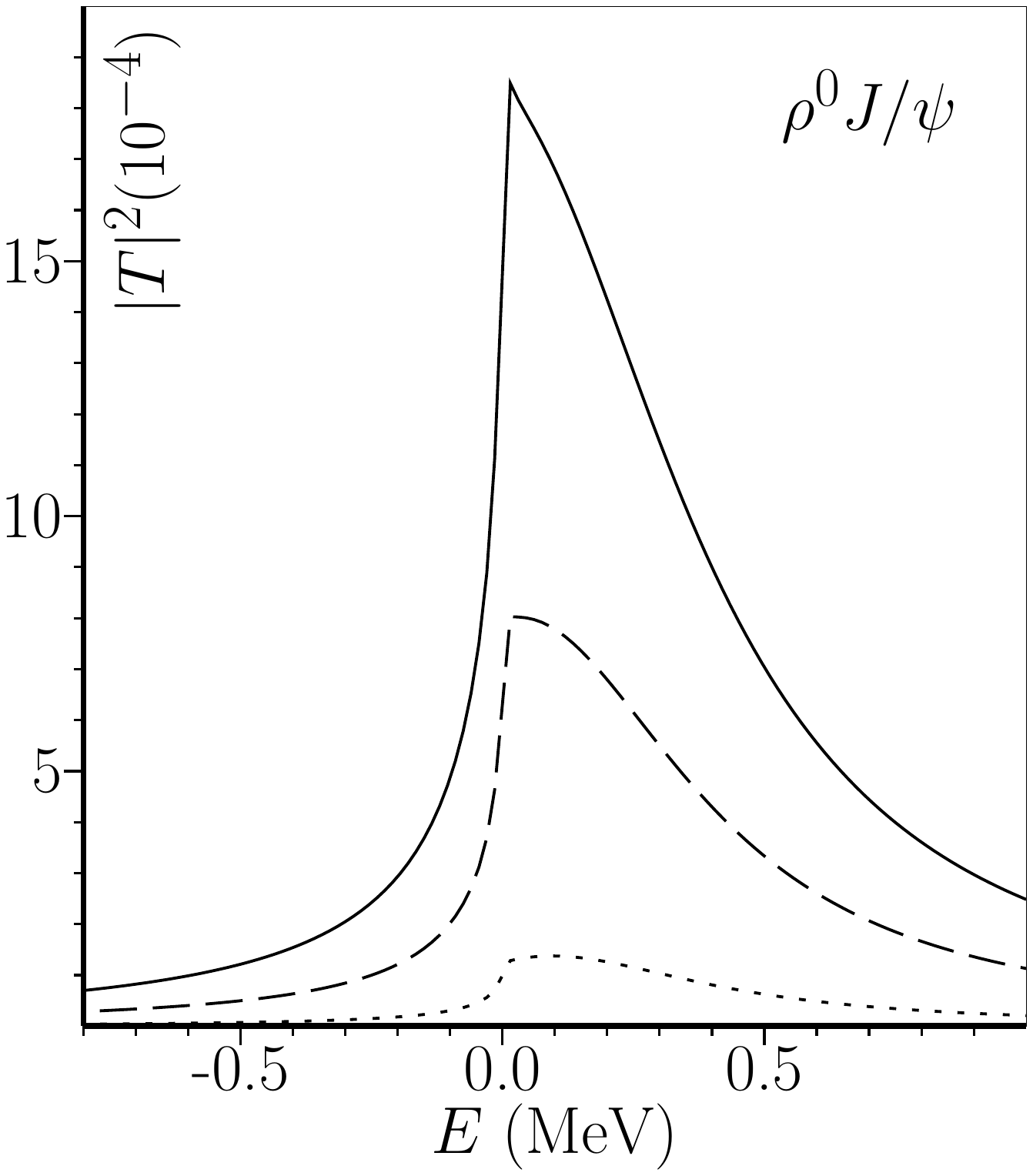}}
\end{tabular}
\mbox{ } \\[-110pt]
\caption{$\rho^0\jpsi$
elastic
amplitude for reduced $\rho^0$ width.
Left: 0\% (dots), 1\% (dashes), 5\% (full); right: 10\% (dots),
50\% (dashes),
100\%
(full). 
Studied case: pole
2
in Table~\ref{poles}.
Also see Fig.~\ref{trajectories}.}
\label{rhowidth}
\end{figure}

Finally, in
order to study the effect of using a complex mass for the $\rho^0$ in the
$\rho^0\jpsi$ channel, we vary the $\rho$ width from 0\% to
100\%
of its PDG \cite{JPG37p075021} value and plot the corresponding amplitudes in
Fig.~\ref{rhowidth}. We see that the maximum $|T|^2$ increases by
almost 3 orders of magnitude
when going from the 0\% case (dotted curve in left-hand plot) to
the
100\%
case (solid curve in right-hand plot).
Furthermore, the 0\% curve only starts out at the central $\rho^0\jpsi$
threshold, of course.
Thus, it becomes clear that no
realistic description of the $\rho^0\jpsi$ channel is possible without
smearing out somehow its threshold, so that its influence kicks in
before the $D^0D^{\ast0}$ channel opens and depletes the signal.
Naturally, similar conclusions apply in principle to the $\omega\jpsi$
channel, though there the effects are less pronounced because of the small
$\omega$ width and the somewhat higher threshold.
These results show that
our unitarisation procedure for complex masses in the asymptotic states
performs as expected
in accounting for thresholds involving resonances.

To conclude our discussion, we should mention that our results are
qualitatively in agreement with those of Danilkin \& Simonov
\cite{PRD81p074027}, in the sense that a single resonance pole originating
from the $2\,{}^{3\!}P_1$ $c\bar{c}$ state is capable of describing the
$X(3872)$ data. However, we disagree with their conclusions on the 
$2\,{}^{3\!}P_0$ state. In an earlier, single-channel description
\cite{PRD74p037501}, we found a resonance at 3946~MeV with a width of 58~MeV,
and we do not believe a detailed multichannel calculation will change these
values dramatically. Thus, the listed $X(3945)$ \cite{JPG37p075021} resonance,
with mass 3916 MeV, width 40 MeV, and positive $C$-parity, appears to be a good
candidate. As for the $2\,{}^{1\!}P_1$ state, the $X(3940)$ \cite{JPG37p075021}
resonance, with mass 3942 MeV, width 37 MeV, and principal decay mode $DD^\ast$,
seems the obvious choice. With the old $Z(3930)$ meanwhile identified as the
$2\,{}^{3\!}P_2$ ($\chi_{c2}(2P)$ \cite{JPG37p075021}) state, we
might
so understand all 4 charmonium states in the range 3.87--3.95~MeV.

\section{Summary and conclusions}
Summarising, we have investigated the $1^{++}$ charmonium scenario for the 
$X(3872)$ resonance, by analyzing in detail the influence of the
$D^0D^{\ast0}$,
$\rho^0\jpsi$, and $\omega\jpsi$
channels on pole positions and amplitudes.
In order to describe the latter OZIF
channels
in a realistic way, we have
used
complex masses for the $\rho^0$ and $\omega$,
and then restored unitarity of the
$S$-matrix by a new
and rigorous algebraic procedure, 
albeit physically heuristic.
It is true that the redefined $S$-matrix may have some unusual analyticity
properties \cite{NPB12p281}, but in our amplitudes no sign was found of any
nearby spurious singularities.
Moreover,
the behaviour of the $\rho^0\jpsi$
amplitude as a function of the $\rho^0$ width gives us confidence in our
approach. Concretely, we have shown that our scenario is compatible with the
$D^0D^{\ast0}$ and $\pi^+\pi^-\jpsi$ data, with a single resonance pole
on top of
or slightly below the $D^0D^{\ast0}$ threshold.
Moreover, our treatment of the $\rho^0\jpsi$ and $\omega\jpsi$ channels
has proven compatible with the observed branching ratio of these decays.

Thus, the data do not
seem to require a molecular or tetraquark interpretation of the $X(3872)$,
also in view of so far unobserved \cite{PRD71p031501} charged partner states.
Nevertheless, only further improved measurements and theoretical calculations
will in the end allow to draw a definitive conclusion on the scenario
preferred by nature.

In conclusion, we must stress that the $X(3872)$, whatever its
assignment, is an extraordinary structure, because of its coincidence ---
to an accuracy of less than 1 MeV --- with the central thresholds of the
principal decay modes. This circumstance is at the same time a blessing and a
curse. To start with the latter, no model can ambition to quantitatively 
describe the $X(3872)$ with present-day state-of-the-art in strong
interactions, while experiment will have an extremely hard time to reduce the
bin sizes to less than 1 MeV and simultaneously keep statistics sufficiently
high. On the other hand, with a strengthened effort of both theory and
experiment, a wealth of knowledge on charmonium spectroscopy and strong decay
--- OZI-allowed as well as OZI-forbidden --- may be gathered by further
studying this fascinating resonance.

\begin{acknowledgement}
We are indebted to Prof.~D.~V.~Bugg and J.~Segovia for very useful comments.
This work was supported by
the \emph{Funda\c{c}\~{a}o para a Ci\^{e}ncia e a Tecnologia}
\/of the \emph{Mi\-nist\'{e}rio da Ci\^{e}ncia, Tecnologia e Ensino Superior}
\/of Portugal, under contracts nos.\ CERN/FP/83502/2008 and
CERN/FP 109307/2009.
\end{acknowledgement}

\appendix

\section{Multichannel $T$-matrix}
\label{tmatrix}
The off-energy-shell RSE $T$-matrix reads \cite{AOP324p1620,PRD80p094011}
\begin{eqnarray}
\hspace*{-15pt}
\lefteqn{\tmat{i}{j}(p_i,p'_j;E)=-2\lambda^2\sqrt{\mu_ip_ir_i}\,\bes{i}(p_ir_i)
\,\times} \nonumber \\
&&\hspace*{-20pt}\sum_{m=1}^{N}\rse_{im}\left\{[\One-\Omega\,\mathcal{R}]^{-1}
\right\}_{\!mj}\bes{j}(p'_jr_j)\,\sqrt{\mu_jp'_jr_j} \; ,
\label{tmat}
\end{eqnarray}
with the diagonal loop function
\begin{equation}
\Omega_{ij}(k_j) =
-2i\lambda^2\mu_jk_jr_j\,\bes{j}(k_jr_j)\,\han{j}(k_jr_j)\,\delta_{ij}\;, 
\label{omega}
\end{equation}
and the RSE propagator
\begin{equation}
\mathcal{R}_{ij}(E)=\sum_{n=0}^{\infty}
\frac{g^i_{(l_c,n)}g^j_{(l_c,n)}}{E-E_n^{(l_c)}}\;.
\label{rse}
\end{equation}
Here, the RSE propagator contains an infinite tower of $s$-channel bare
$q\bar{q}$ states, corresponding to the spectrum of an, in principle,
arbitrary confining potential. Furthermore, $\lambda$ is an overall coupling,
$r_i$ is the decay radius of meson-meson (MM) channel $i$, $E_n^{(l_c)}$ is the
discrete energy of the $n$-th recurrence in the $q\bar{q}$ channel with angular
momentum $l_c$, $g^i_{(l_c,n)}$ is the corresponding coupling to the $i$-th
MM channel, $\mu_i$ the reduced mass for this channel, $p_i$ the off-shell
relative momentum , and $\bes{i}(p_i)$ and $\han{j}(k_jr_j)$ the spherical
Bessel and Hankel functions of the first kind, respectively. Note that $\mu_i$,
$p_i$, and $k_i$ (the on-energy-shell relative momentum) are defined
relativistically (see e.g.\ Ref.~\cite{EPJC22p493}, Eqs.~(16,17)). Also notice
that, for a $J^{PC}=1^{++}$ $c\bar{c}$ system, there is only one confined
channel, viz.\ with $l_c=1$.

Bound states and resonances can be found by searching for zeros in the
determinant of the matrix $(\One\!-\!\Omega\,\mathcal{R})$ in
Eqs.~(\ref{tmat}--\ref{rse}), employing a complex Newton's method. On the
other hand, elastic and inelastic amplitudes are obtained by taking the
on-shell values of the corresponding matrix elements of $T$.

\section{Redefining the $S$-matrix}
\label{unitarization}
It is straightforward to show that the $S$-matrix $S(E)\equiv\One+2iT(E)$,
where $T(E)$ is the on-energy-shell restriction of the multichannel $T$-matrix
in Eqs.~(\ref{tmat}--\ref{rse}), is unitary and symmetric, when limited to open
channels and real energies. However, it is also easy to see 
that, for complex masses and so complex relative momenta, the unitarity of $S$
is lost, though not its symmetry. The latter property can be used to redefine
the physical $S$-matrix.

Since $S$ is always a symmetrix matrix, it can
be decomposed, via Takagi \cite{JJM1p82} factorisation, as
\be
S \; = \; VDV^{T} \; ,
\label{takagi}
\ee
where $V$ is unitary and $D$ is a real nonnegative diagonal matrix.
Then we get
\be
S^\dag S = (V^T)^\dag DV^\dag VDV^T = (V^T)^\dag D^2V^T
= U^\dag D^2U ,
\label{sdags}
\ee
where we have defined $U\equiv V^T$, obviously unitary, too. So
the diagonal elements of $D=\sqrt{US^\dag SU^\dag}$  are the square
roots of the eigenvalues of the positive Hermitian matrix $S^\dag S$,
which are all real and nonnegative. Moreover, since $S=\One+2iT$ is
manifestly nonsingular, the eigenvalues of $S^\dag S$ are even all
nonzero and $U$ is unique. Thus, we may define
\be
S^\prime \; \equiv \; SU^\dag D^{-1}U \; .
\label{sprime}
\ee
Then, using Eq.~(\ref{takagi}) and $V=U^T$, we have
\be
S^\prime \; = \; U^TDUU^\dag D^{-1}U \; = \; U^TU \; ,
\label{sprimesym}
\ee
which is obviously symmetric and,
as
\be
(U^TU)^\dag=U^\dag(U^\dag)^T=U^{-1}(U^{-1})^T=(U^TU)^{-1}\;,
\label{sprimeunit}
\ee
also unitary. So $S^\prime$ has the required properties to be defined as
the $S$-matrix for a scattering process with complex masses in the asymptotic
states.

\end{document}